# Analysis and Implementation of the SNOW 3G Generator Used in 4G/LTE Systems

J. Molina-Gil[1], P. Caballero-Gil[1], C. Caballero-Gil[1], A. Fúster-Sabater[2]

[1]Department of Statistics, O.R. and Computing. University of La Laguna. Spain.
Email:{jmmolina, pcaballe, ccabgil}@ull.es
[2]Institute of Applied Physics. Spanish National Research Council. Madrid, Spain
Email: amparo@iec.csic.es

**Abstract.** The fourth generation of cell phones, marketed as 4G/LTE (Long-Term Evolution) is being quickly adopted worldwide. Given the mobile and wireless nature of the involved communications, security is crucial. This paper includes both a theoretical study and a practical analysis of the SNOW 3G generator, included in such a standard for protecting confidentiality and integrity. From its implementation and performance evaluation in mobile devices, several conclusions about how to improve its efficiency are obtained.

**Keywords:** 4G/LTE encryption, stream cipher, SNOW 3G.

## 1 Introduction

The large increase of mobile data use and the emergence of broadband demanding applications and services are the main motivations for the proposal of progressive substitution of 3G/UMTS by 4G/LTE technology. Nowadays, commercial LTE networks have been launched in many countries. In particular they include: four countries of Africa, between 2012 and 2013; eleven countries of America, including USA from 2010; nineteen countries in Asia, where Japan was the main technology promoter; twenty-nine countries in Europe, excluding Spain even though being one of the largest countries; and two countries in Oceania.

In general, each evolution of telecommunications systems has involved the improvement of security features thanks to the learning from weaknesses and attacks suffered by their predecessors. Regarding the encryption systems used to protect confidentiality in mobile phone conversations, the evolution has been the following. First, the stream cipher A5/1 and its A5/2 version were developed for the 2G/GSM cell phone standard. Serious weaknesses in both ciphers were identified so the encryption system listed in the 3G/UMTS standard substituted them by a completely different scheme, the Kasumi block cipher. In 2010, Kasumi was broken with very modest computational resources. Consequently, again the encryption system was changed in the new standard 4G/LTE, where the stream cipher SNOW 3G is used for protecting confidentiality and integrity.

The main issue of this work is the practical analysis of the SNOW 3G generator, which is the core of both the confidentiality algorithm UEA2 and the integrity algorithm UIA2, published in 2006 by the 3GPP Task Force [1]. Its choice allows higher speed data rates in cell phones thanks to its efficiency when implemented in devices with limited resources. The theoretical analysis of the security level of the SNOW 3G is out of the scope of this paper.

This work is organized as follows. A brief discussion on related work is included in Section 2. Then, Section 3 introduces the main concepts and notations used throughout this work, together with a theoretical description of the SNOW 3G generator. Section 4 gives some details of the implementation carried out in the iPhone Operating System (iOS), and its performance evaluation. Finally, Section 5 closes this paper with some conclusions and future work.

## 2     Related Work

The predecessors of SNOW 3G are SNOW 1.0 [2] and SNOW 2.0 [3].

The original version, SNOW 1.0, was submitted to the NESSIE project, but soon a few attacks were reported. One of the first published attacks was a key recovery requiring a known output sequence of length $2^{95}$, with expected complexity $2^{224}$ [4]. Another cryptanalysis was a distinguishing attack [5], also requiring a known output sequence of length $2^{95}$ and about the same complexity.

Those and other attacks demonstrated some weaknesses in the design of SNOW 1.0, so a more secure version called SNOW 2.0, was proposed. SNOW 2.0 is nowadays one of two stream ciphers chosen for the ISO/IEC standard IS 18033-4 [6]. Also, SNOW 2.0 uses similar design principles to the stream cipher called SOSEMANUK, which is one of the final four Profile 1 (software) ciphers selected for the eSTREAM Portfolio [7].

Afterwards, during its evaluation by the European Telecommunications Standards Institute (ETSI), the design of SNOW 2.0 was further modified to increase its resistance against algebraic attacks [8] with the result named SNOW 3G. Full evaluation of the design of SNOW 3G has not been made public, but a survey of it is given by ETSI in [9].

The designers and external reviewers show that SNOW 3G has remarkable resistance against linear distinguishing attacks [10, 11], but SNOW 3G have suffered other types of attacks. One of the first and simplest cryptanalytic attempts was the fault attack proposed in [12]. An approach to face that problem includes employing nonlinear error detecting codes. A cache-timing attack [13] on SNOW 3G, based on empirical timing data, allows recovering the full cipher state in seconds without the need of any known keystream. Such an attack is based on the fact that operations like the permutations and multiplications by the constant α and its inverse are actually implemented using lookup tables. The work [14] describes a study of the resynchronization mechanism of SNOW 3G using multiset collision attacks, showing a simple 13-round multiset distinguisher with complexity of $2^8$ steps.

The SNOW 3G generator has been subject of a few review works [15, 16]. The present paper provides a new study, more focused on a practical view.

## 3    Theoretical Description of the SNOW 3G Generator

Stream ciphers are based on generators of pseudo-random keystream sequence whose bits are bitwise XORed with the plaintext in order to generate the ciphertext. The main advantage of stream ciphers is that they are lightweight and can operate at a high speed, making them extremely suitable for power-constrained devices such as mobile phones. The stream generator analysed in this work has a typical nonlinear structure based on a Linear Feedback Shift Register (LFSR).

The following terms and notation are used within this paper to describe the stream cipher SNOW 3G and its implementation:

| | |
|---|---|
| $GF(2)=\{0,1\}$ | Galois Field with two elements 0 and 1. |
| $GF(2)[x]$ | Ring of polynomials in the variable x with coefficients in $GF(2)$. |
| d | Degree of a polynomial. |
| p(x) | Primitive polynomial of degree d in $GF(2)[x]$. |
| $GF(2^d)$ | Extension field of $GF(2)$ defined by p(x), with $2^d$ elements. |
| $GF(2^d)[x]$ | Ring of polynomials in the variable x with coefficients in $GF(2^d)$. |
| $\beta \in GF(2^8)$ | Root of the $GF(2)[x]$ polynomial $x^8 + x^7 + x^5 + x^3 + 1$. |
| $\alpha \in GF(2^{32})$ | Root of the $GF(2^8)[x]$ polynomial $x^4 + \beta^{23}x^3 + \beta^{245}x^2 + \beta^{48}x + \beta^{239}$. |
| $s_t$ | 32-bit stage of an LFSR. |
| = | Assignment operator. |
| $\oplus$ | Bitwise XOR operation. |
| $\boxplus$ | Integer addition modulo $2^{32}$. |
| \|\| | Concatenation of two operands. |

As shown in Fig. 1, the SNOW 3G generator consists of two main components: an LFSR and a Finite State Machine (FSM).

The LFSR component has 16 stages $s_0$, $s_1$, $s_2$,..., $s_{15}$, each holding 32 bits. Its feedback is defined by a primitive polynomial over the finite field $GF(2^{32})$, and involves two multiplications, one by a constant $\alpha \in GF(2^{32})$ and another by its inverse, as described by the following relation:

$$s_{t+16} = \alpha\, s_t \oplus s_{t+2} \oplus \alpha^{-1}\, s_{t+11}, \text{ for } t \geq 0. \qquad (1)$$

The FSM component constitutes the nonlinear part of the generator. The FSM involves two input data from the LFSR, which are the $s_5$ and $s_{15}$ stages contents. The FSM is based on three 32-bit registers R1, R2 and R3, and two substitution boxes S1 and S2 that are used to update the registers R2 and R3. Both S-Boxes S1 and S2 map each 32-bit input to a 32-bit output by applying several combinations of a basic S-box on each one of the 4 bytes of the input. However, while box S1 is based on the AES (Advanced Encryption Standard) S-box, the basic S-box of S2 was specially designed for SNOW 3G. The mixing operations in the FSM are bitwise XOR operations and integer additions modulo $2^{32}$.

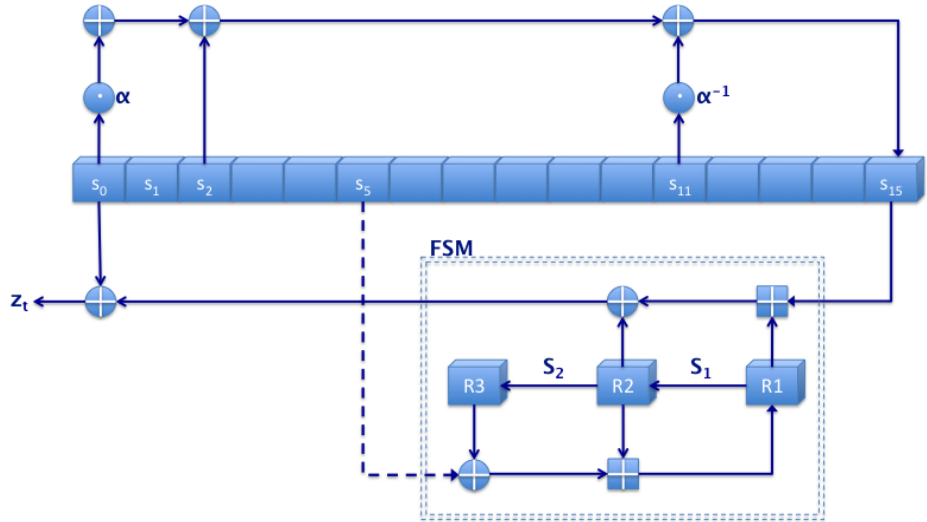

**Fig. 1.** SNOW 3G Generator

The clocking of the LFSR component of SNOW 3G has two different modes of operation, the initialisation mode and the keystream mode. On the one hand, when the initialisation is performed, the generator is clocked without producing any output. On the other hand, in the keystream mode, with every clock tick the generator produces a 32-bit word. Thus, SNOW 3G is a word-oriented generator that outputs a sequence of 32-bit words under the control of a 128-bit key and a 128-bit Initialization Vector IV.

Regarding the implementation of SNOW 3G, which is the main object of the following section, several observations can be done. First, the two multiplications involved in the LFSR can be implemented as a byte shift together with an unconditional XOR with one of $2^8$ possible patterns, as shown below.

Since β is a root of the primitive polynomial $x^8 + x^7 + x^5 + x^3 + 1$, the extension field $GF(2^8)$ can be generated through successive powers of β so that {0, 1, β, $β^2$, $β^3$, …, $β^{2^8-2}$} is the entire field $GF(2^8)$. Thus, any element of $GF(2^8)$ can be represented either with a polynomial in $GF(2)[x]$ of degree less than 8, or with a byte whose bits correspond to the coefficients in such a polynomial. Operations in $GF(2^8)$ correspond to operations with polynomials modulo $x^8 + x^7 + x^5 + x^3 + 1$. This means that, in particular, the multiplication of two elements in $GF(2^8)$ results from the multiplication of the two corresponding polynomials, which is then divided by the polynomial $x^8 + x^7 + x^5 + x^3 + 1$, so that the remainder is the resulting output. The implementation of this operation as a binary multiplication is as follows. Considering both multiplier bytes, for each 1 bit in one of the multipliers, a number of left shifts are run on the other multiplier byte followed, every time the leftmost bit of the original byte before the shift is 1, by a conditional bitwise XOR with $A9_{16}=10101001_2$, which is the byte corresponding to the polynomial $x^8 + x^7 + x^5 + x^3 + 1$. The number of left shifts is given by the position of the 1 bit in the first multiplier.

Since α is a root of the primitive $GF(2^8)[x]$ polynomial $x^4 + \beta^{23}x^3 + \beta^{245}x^2 + \beta^{48}x + \beta^{239}$, the finite extension field $GF(2^{32})$ can be generated through successive powers of α so that $\{0, 1, \alpha, \alpha^2, \alpha^3, ..., \alpha^{2^{32}-2}\}$ is the entire field $GF(2^{32})$. Thus, we can represent any element of $GF(2^{32})$ either with a polynomial in $GF(2^8)[x]$ of degree less than 4, or with a word of 4 bytes corresponding to the 4 coefficients in such a polynomial. Operations in $GF(2^{32})$ correspond to operations with polynomials modulo $x^4 + \beta^{23}x^3 + \beta^{245}x^2 + \beta^{48}x + \beta^{239}$. This means that, in particular, the multiplication of α and any 4-byte word $(c_3, c_2, c_1, c_0)$ in $GF(2^{32})$ results from the multiplication of x and the polynomial $c_3x^3 + c_2x^2 + c_1x + c_0$, which is then divided by the polynomial $x^4 + \beta^{23}x^3 + \beta^{245}x^2 + \beta^{48}x + \beta^{239}$, so that the resulting output is the remainder $(c_2+c_3\beta^{23})x^3 + (c_1+c_3\beta^{245})x^2 + (c_0+c_3\beta^{48})x + c_3\beta^{239}$, or equivalently, the 4-byte word $(c_2+c_3\beta^{23}, c_1+c_3\beta^{245}, c_0+c_3\beta^{48}, c_3\beta^{239})$. Thus, a fast binary implementation of this operation can be based on precomputed tables $(c\beta^{23}, c\beta^{245}, c\beta^{48}, c\beta^{239})$, $\forall c \in GF(2^8)$. Similarly, the multiplication of $\alpha^{-1}$ and any 4-byte word $(c_3, c_2, c_1, c_0)$ in $GF(2^{32})$ results from the multiplication of $x^{-1}$ and the polynomial $c_3x^3 + c_2x^2 + c_1x + c_0$, which is $c_3x^2 + c_2x + c_1 + c_0 x^{-1}$. Since $xx^{-1}=1$ and $\beta^{255}=1$, $x^{-1}$ can be expressed as $\beta^{255-239}x^3 + \beta^{255-239+23}x^2 + \beta^{255-239+245}x + \beta^{255-239+48} = \beta^{16}x^3 + \beta^{39}x^2 + \beta^6x + \beta^{64}$. Thus, the resulting output of the product is the remainder $(c_0\beta^{16})x^3+(c_3+c_0\beta^{39})x^2+(c_2+c_0\beta^6)x+(c_1+c_0\beta^{64})$, or equivalently, the 4-byte word $(c_0\beta^{16}, c_3+c_0\beta^{39}, c_2+c_0\beta^6, c_1+c_0\beta^{64})$. Thus, a fast binary implementation of this operation can be based on precomputed tables $(c\beta^{16}, c\beta^{39}, c\beta^6, c\beta^{64})$, $\forall c \in GF(2^8)$.

## 4   iOS Implementation and Evaluation

This work analyses a software implementation of SNOW 3G in cell phone platform. In particular, we have implemented it for iOS platform and the used programming language has been Objective C.

The first aspect we have taken into account is that LFSRs have been traditionally designed to operate over the binary Galois field $GF(2)$. This approach is appropriate for hardware implementations but its software efficiency is quite low. Since microprocessors of most cell phones have a word length of 32 bits, the LFSR implementation is expected to be more efficient for extended fields $GF(2^{32})$. Thus, since the implementation of SNOW 3G is over the finite field $GF(2^{32})$, it is more suitable for the architecture that supports current cell phones. The second aspect is related to arithmetic operations and specifically, the multiplication on extension fields of $GF(2)$ because the feedback function in SNOW 3G involves several additions and multiplications, and the multiplication is the most computationally expensive operation.

In this section, we study and compare different software implementation in order to find the optimal one for devices with limited resources, such as smartphones. We have performed several studies on an iPhone 3GS whose main characteristics are described in Table 1.

**Table 1.** Device used for the evaluation

| iPhone 3GS | | | |
|---|---|---|---|
| *Architecture* | *CPU Frequency* | *Cache L1I/L1D/L2* | *RAM* |

| Armv7-A | 600 MHz | 16 Kb/16 Kb/256 Kb | 256 MB |

All the results shown in this work have been obtained using Instruments, which is a tool for analysis and testing of performance of OS X and iOS code. It is a flexible and powerful tool that lets track one or more processes and examine the collected data. The tests correspond to the average of 10 runs in which $10^7$ bytes of keystream sequence are generated using the platform described above. Table 2 shows the total time (in milliseconds) for each SNOW 3G function explained below. The evidences indicate that the multiplication is the most expensive function. The second most expensive function is the shift register, which is performed in each clock pulse.

Below we study two different techniques to perform multiplications and several techniques for LFSR software implementation proposed in [11].

**Table 2.** Function Performance in Recursive Mode

| Summary | | |
|---|---|---|
| Function | *Time(ms)* | *%* |
| MULxPow | 29054,9 | 92,88 |
| ClockLFSRKeyStreamMode | 572 | 1,77 |
| DIValpha | 356,6 | 1,1 |
| main | 264,7 | 0,8 |
| MULalpha | 326,8 | 0,99 |
| GenerateKeystream | 243,8 | 0,73 |
| ClockFSM | 180,3 | 0,54 |
| S2 | 128,1 | 0,34 |
| S1 | 129,9 | 0,37 |
| Generator | 1,3 | 0 |
| **Total Time** | 30258,5 | |

### 4.1 Multiplication

As shown in Table 2, according to the implementation proposed in [1] the most consuming time function in SNOW 3G is the MULxPow used in both the multiplication by α and by $α^{-1}$. Each multiplication can be implemented either as a series of recursive byte shifts plus additional XORs, or as a lookup table with precomputed results. In each clocking of the LFSR, the feedback polynomial uses two functions $MUL_α$ and $DIV_α$ which are defined as:

$$MUL_α = MUL_XPOW(c, 23, 0xA9) \,||\, MUL_XPOW(c, 245, 0xA9)$$
$$MUL_XPOW(c, 48, 0xA9) \,||\, MUL_XPOW(c, 239, 0xA9)$$

$$DIV_α = MUL_XPOW(c, 16, 0xA9) \,||\, MUL_XPOW(c, 39, 0xA9)$$
$$MUL_XPOW(c, 6, 0xA9) \,||\, MUL_XPOW(c, 64, 0xA9)$$

The first method might be more appropriate for systems with limited memory resources, as it does not require a large storage. However, as we can see in Table 2, it has a significant computational cost.

The second method involving precomputed tables provides optimal time results, as can be seen in Table 3. Indeed, it can be considered the fastest procedure for multiplication because it results in an improvement of 96% in time consumption with respect to the first recursive method. However, one of the biggest problems with this proposal could be the needed storage in devices with limited resources. In particular, for SNOW 3G, the table has 256 elements, each of 32 bits, what results in a total of 32*256 bits. Furthermore, the implementation uses the two functions MUL$\alpha$ and DIV$\alpha$, so it involves two tables, what means a total of 2048 bytes. Consequently, this method seems quite adequate for the characteristics of the chosen device.

**Table 3.** Function Performance With precomputed tables

| Computational Cost | | |
| --- | --- | --- |
| Function | *Time(ms)* | *%* |
| ClockLFSRKeyStreamMode | 347,3 | 28,69 |
| main | 277,2 | 22,35 |
| ClockFSM | 182,2 | 14,95 |
| S1 | 146 | 12,01 |
| S2 | 138 | 11,3 |
| GenerateKeystream | 107,3 | 8,84 |
| Generator | 1,3 | 0,04 |
| **Total Time** | 1199,4 | |

### 4.2 LFSR

The LFSR structures are difficult to implement efficiently in software. The main reason is the shift of each position during each clock pulse. This shift in hardware implementation occurs simultaneously, so the whole process can be performed in a single clock pulse. However, in software implementation, the process is iterative and costly.

As we saw in Table 3, once optimized the multiplication, it is the ClockLFSRKeyStreamMode function the most time consuming. Thus, we have used different software optimization techniques, proposed in [17] together with the hardcode technique presented in the specifications in order to improve the LFSR's final performance.

The hardcode method consists in embedding the data directly into the source code, instead of using loops or indices as the rest of techniques do. The cost of this proposal corresponds to 15 assignments. This technique, despite being longer, seems to require less time. Below is the implementation of this method.

```
void ClockLFSRKeyStreamMode()
{
 u32 v = ( ( (LFSR_S0 << 8) & 0xffffff00 ) ^
 ( MULalpha( (u8)((LFSR_S0>>24) & 0xff) ) ) ^
```

```
        ( LFSR_S2 ) ^
        ( (LFSR_S11 >> 8) & 0x00ffffff ) ^
        ( DIValpha( (u8)( ( LFSR_S11) & 0xff ) ) )
    );
    LFSR_S0 = LFSR_S1;
    LFSR_S1 = LFSR_S2;
    LFSR_S2 = LFSR_S3;
    LFSR_S3 = LFSR_S4;
    LFSR_S4 = LFSR_S5;
    LFSR_S5 = LFSR_S6;
    LFSR_S6 = LFSR_S7;
    LFSR_S7 = LFSR_S8;
    LFSR_S8 = LFSR_S9;
    LFSR_S9 = LFSR_S10;
    LFSR_S10 = LFSR_S11;
    LFSR_S11 = LFSR_S12;
    LFSR_S12 = LFSR_S13;
    LFSR_S13 = LFSR_S14;
    LFSR_S14 = LFSR_S15;
    LFSR_S15 = v;
}
```

The analysis carried out with the precomputed multiplication method involves an experiment to assess $10^7$ bytes of the keystream generated by the LFSR proposed for SNOW 3G. The result values are summarized in Table 4, which shows the functions' time and the total implementation time.

The results show that the hardcode method is not the best implementation. Although it represents an 11% of improvement over the traditional method, the sliding windows method presents an improvement of 29% with respect to the traditional, and 20% compared to the hardcode method.

**Table 4.** Performance of Different LFSR Implementation Methods

|  | *Traditional* | *HardCode* | *Circular Buffers* | *Sliding Windows* | *Loop Unrolling* |
|---|---|---|---|---|---|
| **Function** | *Time (ms)* | *Time (ms)* | *Time (ms)* | *Time (ms)* | *Time (ms)* |
| ClockLFSRKeyStreamMode | 491,1 | 342,5 | 834 | 184,1 | 291,3 |
| Generator | 1,4 | 1,3 | 1,3 | 1,8 | 1,1 |
| GenerateKeystream | 65,8 | 65,8 | 198,3 | 88,2 | 68,4 |
| main | 246,6 | 306,8 | 296,8 | 297 | 294,4 |
| **Total Time** | 804,9 | 716,4 | 1330,4 | **571,1** | 655,2 |

From the obtained results, we conclude that the circular buffer method is not applicable because the update of different indices involves modular arithmetic, which is not very efficient.

The new LFSR proposal can affect other SNOW 3G parts like the FSM. For this reason our main aim is to determine whether improving LFSR shift times negatively affects other code parts, and in that case to state the improvement that can be achieved. In order to do it, we have implemented in SNOW 3G with precomputed tables using sliding windows for shift register, Table 5 shows the summary of the results. If we compare them with the results of Table 3, it is clear that this implementation improves the time for the *ClockLFSRKeyStreamMode*, *S1* and *GenerateKey*. However, other functions like *ClockFSM*, *S2, GenerateKeystream* have increased slightly their values. The function with the worst time result is *S2,* as its value has increase 26% related to the previous proposal. Moreover, the greatest improvement has been in *ClockLFSRKeyStreamMode* function, with a 47%. All this results in an overall improvement of 10% compared to the implementation proposed in the specifications.

**Table 5.** Function Performance in optimized Mode

| Computational Cost | | |
|---|---|---|
| **Function** | *Time(ms)* | *%* |
| ClockLFSRKeyStreamMode | 184,7 | 15,28 |
| main | 282,3 | 22,23 |
| ClockFSM | 195,4 | 17,68 |
| S1 | 135,9 | 12,65 |
| S2 | 163,4 | 16,33 |
| GenerateKeystream | 118,2 | 10,95 |
| Generator | 1,2 | 1,07 |
| **Total Time** | 1081,1 | |

## 5    Conclusions and Future Work

This paper has provided an analysis both from a theoretical and practical point of view, of the generator used for the protection of confidentiality and integrity in the 4G/LTE generation of mobile phones. In particular, after an introduction to the theoretical basis of the SNOW 3G generator, the implementation of the generator on the iOS mobile platform and several experiments have been carried out, obtaining from a comparison with similar software, several interesting conclusions on how to improve efficiency through the optimization of the software. Since this is an on-going work, there are still many open problems such as the analysis of other parameters not yet analyzed in this work, the implementation using different architectures and a comparative study. Also other future works are the proposal of a lightweight version of the SNOW 3G generator for devices with limited resources, and the analysis of several theoretical properties of the generator.

## Acknowledgements

Research supported by Spanish MINECO and European FEDER Funds under projects TIN2011-25452 and IPT-2012-0585-370000, and the FPI scholarship BES-2009-016774.